\def\year{2022}\relax
\documentclass[letterpaper]{article} 
\usepackage{aaai22}  
\usepackage{times}  
\usepackage{helvet} 
\usepackage{courier}  
\usepackage[hyphens]{url}  
\usepackage{graphicx} 
\usepackage{natbib}  
\usepackage{caption} 

\usepackage{booktabs}
\usepackage{array}
\newcommand{\PreserveBackslash}[1]{\let\temp=\\#1\let\\=\temp}
\newcolumntype{C}[1]{>{\PreserveBackslash\centering}p{#1}}
\newcolumntype{R}[1]{>{\PreserveBackslash\raggedleft}p{#1}}
\newcolumntype{L}[1]{>{\PreserveBackslash\raggedright}p{#1}}
\usepackage{enumerate}
\usepackage{amsmath}
\usepackage{mathtools}
\usepackage{graphicx}
\DeclareGraphicsExtensions{.eps,.ps,.jpg,.bmp}
\usepackage{lineno}
\usepackage{subfigure}
\usepackage{amssymb}
\usepackage{multirow}
\usepackage{algorithm}
\usepackage[noend]{algorithmic}
\usepackage{ifpdf}
\usepackage{verbatim}

\title{GS$^2$-RS: Generating Self-Serendipity Preference in Recommender Systems for Addressing Cold Start Problems}
\author{Yuanbo Xu, Yongjian Yang, En Wang}
\affiliations{

}

\begin{document}

\maketitle

\begin{abstract}
Classical accuracy-oriented Recommender Systems (RSs) typically face the cold-start problem and the filter-bubble problem when users suffer the familiar, repeated, and even predictable recommendations, making them boring and unsatisfied. To address the above issues, serendipity-oriented RSs are proposed to recommend appealing and valuable items significantly deviating from users' historical interactions and thus satisfying them by introducing unexplored but relevant candidate items to them. In this paper, we devise a novel serendipity-oriented recommender system (\textbf{G}enerative \textbf{S}elf-\textbf{S}erendipity \textbf{R}ecommender \textbf{S}ystem, \textbf{GS$^2$-RS}) that generates users' self-serendipity preferences to enhance the recommendation performance. Specifically, this model extracts users' interest and satisfaction preferences, generates virtual but convincible neighbors' preferences from themselves, and achieves their self-serendipity preference. Then these preferences are injected into the rating matrix as additional information for RS models. Note that GS$^2$-RS can not only tackle the cold-start problem but also provides diverse but relevant recommendations to relieve the filter-bubble problem. Extensive experiments on benchmark datasets illustrate that the proposed GS$^2$-RS model can significantly outperform the state-of-the-art baseline approaches in serendipity measures with a stable accuracy performance.  
\end{abstract}
\section{Introduction}
Recent decades have witnessed the magnificent success of recommender systems in both industry and academia. As an important tool to filter the enormous information, a proper recommender system aims to select relevant candidate items for object users while extracting their personalized preferences from their historical shopping logs simultaneously. To achieve this goal, collaborative filtering (CF) \cite{A2DBLP:conf/sigir/ZouXGZ0HY20} and matrix factorization (MF) \cite{A5DBLP:conf/aaai/Chen0ZSCF020} have been the most popular algorithms and were developed for decades. However, as shown in the recent research literature, conventional CF and MF models often suffer from cold-start situations \cite{A3DBLP:conf/sigir/ChaeKCK20}, filter bubbles \cite{A6DBLP:conf/recsys/KapoorKTKS15}, and overfitting \cite{A1DBLP:conf/icml/FeldmanFH19}, which results in inaccurate, irrelevant, and repeated recommendations. To make things worse, these recommendations might irritate users and hurt their shopping experience.

	\begin{figure}[tbp]
	\centering
	\includegraphics[width=0.85\columnwidth]{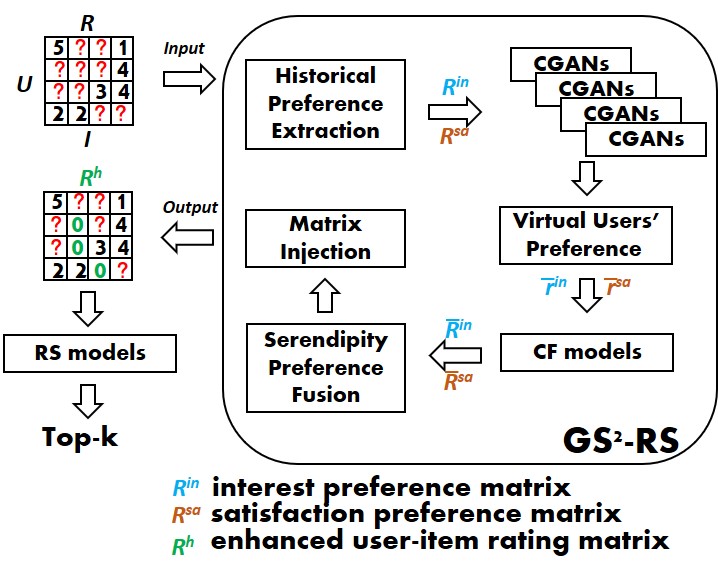}
	\caption{Illustration of our proposed GS$^2$-RS framework for recommendations.}
	\label{SSRS-1}
	\vspace{-10pt}
\end{figure}

In terms of the data sparsity (the sparsity of the user-item interaction matrix), the \textit{Cold-Start} (CS) problem and the \textit{Filter-Bubble} (FB) problem are the vital issues that limit the performance of existing recommendation models. CS problem is a common but terrible challenge for recommender systems. E-commerce websites, such as Amazon, Yelp, and Taobao, usually have millions of users and items. However, the feedback interactions (clicks, browses, purchases, and ratings, etc. In this paper, we focus on ratings.) between them only account for a minimal fraction. Obviously, without enough historical interactions, it is difficult for CF models to understand users' preferences who do not have enough ratings, thereby leading to an inaccurate recommendation. To relieve this cold-start situation, some existing recommendation models prefer to recommend the most popular items to those cold-start users to achieve a better recommendation accuracy expectation. Nevertheless, the similar, repeated recommendations to most users without personalizations result in another critical challenge-the filter-bubble problem-in recommender systems, especially for MF models. With these homogenous recommendations, the popular items' interaction weights increase in MF models. Naturally, those items have more chances to be explored by users than the other items should, dominating the recommendation results, which is a typical scenario of the \textit{Matthew Effect}. What is worse, the CS problem and FB problem usually appear in pairs and reinforce mutually, which seriously impact the recommendation quality.

To address the filter-bubble problem, most researchers so far have focused on exploiting auxiliary information such as users' attributes \cite{A9DBLP:conf/sigir/BiSYWWX20a}, user's social relations \cite{A8DBLP:conf/www/FuZHDHC21}, and item's description text \cite{A3DBLP:conf/sigir/ChaeKCK20} or reviews \cite{A7DBLP:conf/www/0012OM21}. Along with this line, serendipity-oriented recommender system are proposed to make unexpected but valuable items to users \cite{A16DBLP:journals/tmm/YangXWHY18, A17DBLP:journals/jcst/ZiaraniR21}. However, their models are effective and useful only when these auxiliary information are available. Besides, over-utilizing these auxiliary information may cause the privacy disclosure problem \cite{A11DBLP:conf/recsys/BurbachNPZV18}. For the cold-start problem, some studies employed novel data framework to develop the existing recommendation models, such as multi-layer perceptron \cite{A12DBLP:conf/www/HeLZNHC17}, recurrent neural networks \cite{A14DBLP:journals/isci/XuYHWMX19} or the latest graph convolutional network \cite{A13DBLP:conf/sigir/ZhangYCHHC20}, to find the latent neighbors or the similar latent representations of cold-start users/items. However, these neural network-based models require huge computing ability and costly to be deployed. Our research focuses on \textit{original rating-based RS}, a Top-K recommendation task with only one user-item rating matrix, \textit{without requiring any auxiliary information or changing the existing recommendation models}. In this context, \textit{data completion} has been the most popular algorithm to tackle the CS problem \cite{A3DBLP:conf/sigir/ChaeKCK20,A15DBLP:conf/recsys/KimS19,A17DBLP:journals/jcst/ZiaraniR21}. These models fill the sparse user-item rating matrix by inferring the data distribution from existing ratings to relieve the cold-start problem. Nevertheless, data completion models are often too coarse to give a reasonable recommendation (for example, they can not extract users' preferences on items). Thus they can not address the filtering bubble problem. 

In this paper, we propose a novel recommender system framework that can \textit{tackle the cold-start problem and the filter-bubble problem at the same time}. Different from existing neural network-based models or data completion models, our core idea is to generate \textit{virtual but convincible users' preferences on items}, drop out the impossible items from the candidate item set, and enhance most existing recommendation models. Specifically, \textbf{GS$^2$-RS}, which stands for \textbf{G}enerative \textbf{S}elf-\textbf{S}erendipity \textbf{R}ecommendation \textbf{S}ystem, consists of three following modules: 1) preferences modelling. We analyze the user-item rating matrix to form user's historical interest and satisfaction preferences. Then we train different Conditional Generative Adversarial Nets (CGCN) to generate users' virtual preferences (interest and satisfaction) on candidate items. 2) self-serendipity fusion and matrix injection. We devise a gate mechanism to combine users' interests and satisfactions to form their self-serendipity preferences. Then GS$^2$-RS drops out the impossible items from the candidate item set by filling 0s in the user-item rating matrix, which builds an enhanced rating matrix. 3) recommendation. GS$^2$-RS outputs this enhanced rating matrix to any existing recommendation models. Our proposed model can achieve a personalized, customized recommendation by tuning the gate threshold in the matrix injection stage. 

To the best of our knowledge, our work is the first attempt to employ GANs for building a serendipity-oriented recommender system that tackles both CS and FB problems at the same time (as shown in Figure \ref{SSRS-1}). Notably, the contributions of our proposed model are summarized as follows. (i) We propose a novel serendipity-oriented recommendation framework (GS$^2$-RS), which can tackle both cold-start and filter-bubble problems without any auxiliary information. (ii) GS$^2$-RS utilizes GANs to generate users' serendipity preferences from only the user-item rating matrix and makes an explainable, personalized recommendation with a delicate matrix injection method. Specifically, GS$^2$-RS can be treated as the preprocessing for any existing recommendation models. (iii) We conduct extensive empirical studies on three public datasets, and find that GS$^2$-RS can achieve a superior recommendation performance on both accuracy metric and serendipity metric. 

\section{GS$^2$-RS:Generative Self-Serendipity Recommender System}
\subsection{Preliminary and Problem Statement} Given a recommender system, which contains $M$ users and $N$ items, the user-item rating matrix $R$ is a $ M\times N $ low-rank sparse matrix, where its entity $r_{ij}$ stands for the rating that user $i$ marked for item $j$, ranging from [\textit{1,2,3,4,5}]. Note that in real-world scenarios, $R$ is usually extremely sparse (more than 90\% data is unknown to learning models), meaning there are many ``?" in $R$.

\begin{figure}[htbp]
 	\centering
 	\includegraphics[width=0.85\columnwidth]{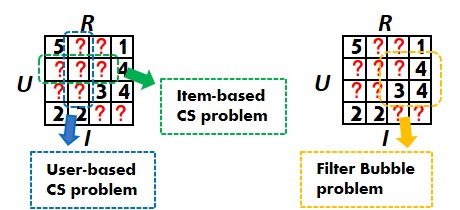}
 	\caption{A simple example to explain CS and FB problem directly.}
 	\label{SSRS-2}
 	\vspace{-10pt}
 \end{figure}

We focus on \textit{original rating-based RS}, which means that for achieving a Top-K recommendation task, the input of our model is only the original user-item rating matrix $R$. However, the sparsity of $R$ usually leads to the cold-start (CS) problem and the filter-bubble (FB) problem \cite{A3DBLP:conf/sigir/ChaeKCK20}, as shown in Figure \ref{SSRS-2}. \textbf{Cold-start problem}: The extreme sparsity may confuse the RS models to infer inaccurate preferences of users on candidate items, which causes item- and user-based CS problems. \textbf{Filter-bubble problem}: Some models pay so much attention to existing ratings that the more ratings an item has, the more chances it can be recommended to users. In Figure \ref{SSRS-2}, as a FB problem, the item with rating [\textit{1,4,4}] dominate other items to be recommended in some existing models, like CF models \cite{A4DBLP:conf/aaai/ChenZZMLM20}. Both the problems affect recommender systems' performance and affect users' shopping experience.
 
\begin{figure}[htbp]
	\centering
	\includegraphics[width=1\columnwidth]{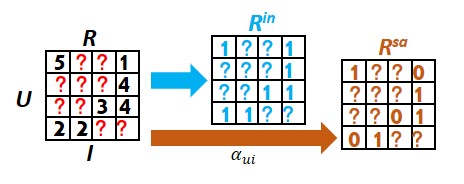}
	\caption{Historical preference extraction. $\alpha_{ui}$ is the $u$'s average rating. Note that for $R^{\text{sa}}$'s fourth row, we take both $u$'s and $i$'s average ratings ((2+2)/2=2; (5+2)/2=3.5) into consideration as the thresholds.}
	\label{SSRS-3}
\end{figure}

To tackle the problem above, we are inspired by \cite{A3DBLP:conf/sigir/ChaeKCK20}, which utilizes GANs to generate users' virtual neighbors for enhancing CF models. However, this model can not consider the FB problem because it directly generates the virtual neighbors' rating without inferring users' preferences. Along with this line, we first introduce users' two type perferences: \textit{interest and satisfaction preferences}: \textit{Interest}: for a user $u$ and an item $i$ in a candidate itemset $i \in I_u^{\text{can}}$, if $r_{ui} \neq ?$, which means that $u$ has the interest to buy $i$, we define user' interest preference $r^{\text{in}}_{ui}=1$, $r^{\text{in}}_{ui} \in R^{\text{in}}$, $i \in I_u^{\text{in}}$. \textit{Satisfaction}: for a user $u$ and an item $i \in I_u^{\text{in}}$, if $r_{ui} \geq \alpha_{ui}$, we define user's satisfaction preference $r^{\text{sa}}_{ui}=1$, $i \in I_u^{\text{sa}}$; else $r^{\text{sa}}_{ui}=0$, $r^{\text{sa}}_{ui} \in R^{\text{sa}}$. Note that the threshold $\alpha_{ui}$ can be defined as either $u$ or $i$'s average rating, or other contextual values. In our proposed model, we first extract users' historical preferences, as shown in Figure \ref{SSRS-3}. 

Meanwhile, we analyze users' preferences at a fine-grained level. We introduce users' \textit{serendipity items}: the items with high relevance but low shopping purpose \cite{A16DBLP:journals/tmm/YangXWHY18}. For common situations, these items can produce a wonderful purchasing experience once the users buy them. So recommend serendipity items can achieve better diversity and satisfaction. By considering serendipity items, the filter-bubble problem can be relieved effectively.

\subsection{Multiple Preference Modelling}
This section introduces how we deduce users' multiple preferences, including users' virtual interest and satisfaction preferences, based on Conditional Generative Adversarial Nets
(CGAN) \cite{A20DBLP:journals/corr/MirzaO14}, a framework to train generative models with complicated, high-dimensional real-world data such as images. Specifically, CGAN is an extension to the original GAN \cite{A19DBLP:conf/nips/GoodfellowPMXWOCB14}: it allows a generative model $\mathcal{G}$ to produce date according to a specific condition vector \textbf{c} by treating the desired condition vector as an additional input with the random noise input \textbf{z}. Thus, CGAN's objective function is formulated as follows:
\begin{equation}
\begin{aligned}
V(\mathcal{D,G}) = {\mathbb{E}_{\textbf{x}-{p_{\text{data}}}(\textbf{x})}}[\ln\mathcal{D}(\textbf{x}|\textbf{c})]-{\mathbb{E}_{\textbf{z}-{p_\textbf{z}}(\textbf{z})}}[\ln\mathcal{D}(\mathcal{G}(\textbf{z}|\textbf{c}))],
\end{aligned}
\label{eq1} 
\end{equation}
where \textbf{x} is a ground truth data from the data distribution $p_{\text{data}}$, $z$ is a noise input vector sampled from known prior $p_\textbf{z}$. $\mathcal{G}(\textbf{z})$ is a synthetic data from the generator distribution $p_\textbf{g}$. $c$ corresponds to a condition vector such as a one-hot vector of a specifc class label. With the optimal objective $\mathop {\min }\limits_{\mathcal{G}} \mathop {\max }\limits_{\mathcal{D}} V(\mathcal{D,G})$. Ideally, the completely trained $\mathcal{G}$ is expected to generate realistic data that the discriminative model $\mathcal{D}$ evaluates the possibility that the data came from the ground truth rather than from $\mathcal{G}$ equals 0.5.

\begin{figure}[tbp]
	\centering
	\includegraphics[width=1\columnwidth]{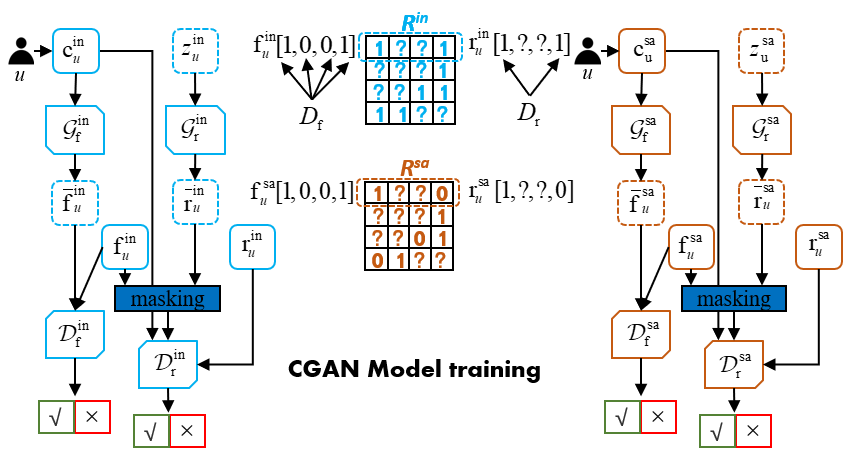}
	\caption{CGAN model training.}
	\label{SSRS-4}
\end{figure}
After we extract users' interest and satisfaction preferences $R^{\text{in}}, R^{\text{sa}}$, we build two CGANs for each type of preferences, respectively. The training procedure is shown in Figure \ref{SSRS-4}. Note that we apply the same CGAN framework on interest and satisfaction preferences, respectively. For the sake of simplicity, we take interest preferences as the example and briefly introduce how this framework can be properly deployed on satisfaction preferences. Formally, we train our two CGANs as follows:
\begin{equation}
\begin{array}{l}
	V({\cal D}_{\rm{r}}^{{\rm{in}}},{\cal G}_{\rm{r}}^{{\rm{in}}},{\cal D}_{\rm{f}}^{{\rm{in}}},{\cal G}_{\rm{f}}^{{\rm{in}}})\\
	\\
	\simeq \frac{1}{{\left| U \right|}}(\sum\limits_{u \in U} {(\ln {\cal D}_{\rm{r}}^{{\rm{in}}}({\bf{r}}_u^{{\rm{in}}}|{\bf{c}}_u^{{\rm{in}}}) - \ln {\cal D}_{\rm{r}}^{{\rm{in}}}({\cal G}_{\rm{r}}^{{\rm{in}}}({\bf{z}}_u^{{\rm{in}}}|{\bf{c}}_u^{{\rm{in}}}) \bullet {\rm{f}}_u^{{\rm{in}}}))} )\\
	{\rm{ + }}\frac{1}{{\left| U \right|}}(\sum\limits_{u \in U} {(\ln {\cal D}_{\rm{f}}^{{\rm{in}}}({\bf{f}}_u^{{\rm{in}}}|{\bf{c}}_u^{{\rm{in}}}) - \ln {\cal D}_{\rm{f}}^{{\rm{in}}}({\cal G}_{\rm{f}}^{{\rm{in}}}({\bf{z}}_u^{{\rm{in}}}|{\bf{c}}_u^{{\rm{in}}})))} ),
\end{array}
\label{eq2}
\end{equation}
where $\textbf{c}_u^{\text{in}}$ denotes $u$'s interest condition vector, $\textbf{z}_u^{\text{in}}$ denotes $u$'s noise vector, $\textbf{r}_u^{\text{in}}$ denotes $u$'s interest vector, $\textbf{f}_u^{\text{in}}$ denotes the interest indicator vector. Note that $\textbf{f}_u^{\text{in}}$ is in the same dimension as $\textbf{r}_u^{\text{in}}$, and each entity $f_{ui}^{\text{in}}$ of $\textbf{f}_u^{\text{in}}$ is $1/0$ to indicate that there is/not an interest value for item $i$. 

In Formula \ref{eq2}, $\mathcal{G}_\text{r}^{\text{in}}$ and $\mathcal{G}_\text{f}^{\text{in}}$ are employed to produce $u$'s synthetic interest vector and interest indicator vector, denoted as $\overline {\bf{r}}_u^{{\rm{in}}}$, $\overline {\bf{f}} _u^{{\rm{in}}}$. While $\mathcal{D}_\text{r}^{\text{in}}$ and $\mathcal{D}_\text{f}^{\text{in}}$ are employed to distinguish real interest vector ${\bf{r}}_u^{{\rm{in}}}$, indicator vector ${\bf{f}}_u^{{\rm{in}}}$ from synthetic vectors $\overline {\bf{r}} _u^{{\rm{in}}}$, $\overline {\bf{f}}_u^{{\rm{in}}}$, respectively. Specifically, there are two designs in this framework: first, each $\mathcal{G}$'s outputs' value are restricted to the range of (0~1) for computing with a Layer Normalization; second, a masking operation is employed to avoid useless computations: ${({\bf{z}}_u^{{\rm{in}}}|{\bf{c}}_u^{{\rm{in}}}) \bullet {\rm{f}}_u^{{\rm{in}}}}$. This element-wise dot operation forces the discriminators to focus on the values $r^{\text{in}}_{ui}$ in interest vector ${\bf{r}}_u^{{\rm{in}}}$, which contributes to the computing results. Besides, the masking operation relieves the data sparsity issue. All $\mathcal{D},\mathcal{G}$ are co-trained and deployed by DNN where the stochastic gradient descent (SGD) with minibatch and the back-propagation algorithm are employed. 

\begin{figure*}[thbp]
	\centering
	\includegraphics[width=1.6\columnwidth]{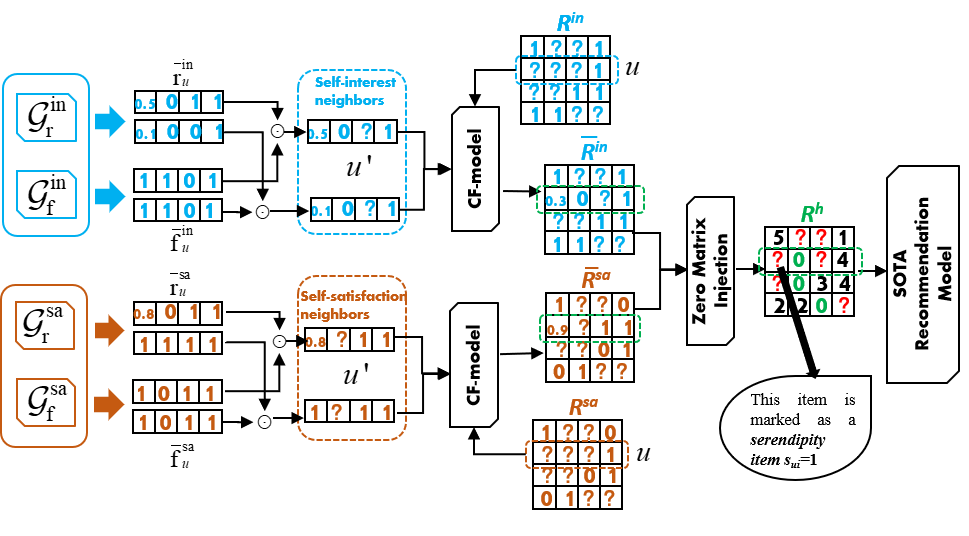}
	\caption{Usage of GS$^2$-RS for enhancing SOTA recommender systems.}
	\label{SSRS-5}
\end{figure*}

Similar as interest preferences, satisfaction preferences can be modeled as follows:
\begin{equation}
\begin{array}{l}
	V({\cal D}_{\rm{r}}^{{\rm{sa}}},{\cal G}_{\rm{r}}^{{\rm{sa}}},{\cal D}_{\rm{f}}^{{\rm{sa}}},{\cal G}_{\rm{f}}^{{\rm{sa}}})\\
	\\
	\simeq \frac{1}{{\left| U \right|}}(\sum\limits_{u \in U} {(\ln {\cal D}_{\rm{r}}^{{\rm{sa}}}({\bf{r}}_u^{{\rm{sa}}}|{\bf{c}}_u^{{\rm{sa}}}) - \ln {\cal D}_{\rm{r}}^{{\rm{sa}}}({\cal G}_{\rm{r}}^{{\rm{sa}}}({\bf{z}}_u^{{\rm{sa}}}|{\bf{c}}_u^{{\rm{sa}}}) \bullet {\rm{f}}_u^{{\rm{sa}}}))} )\\
	{\rm{ + }}\frac{1}{{\left| U \right|}}(\sum\limits_{u \in U} {(\ln {\cal D}_{\rm{f}}^{{\rm{sa}}}({\bf{f}}_u^{{\rm{sa}}}|{\bf{c}}_u^{{\rm{sa}}}) - \ln {\cal D}_{\rm{f}}^{{\rm{sa}}}({\cal G}_{\rm{f}}^{{\rm{sa}}}({\bf{z}}_u^{{\rm{sa}}}|{\bf{c}}_u^{{\rm{sa}}})))} ),
\end{array}
\label{eq3}
\end{equation}

After the training of our CGANS, we are ready to generate users' virtual preferences, including interest and satisfaction. With these virtual preferences, we aim at deducing users' self-serendipity preferences for an accurate recommendation. 

\subsection{Self-serendipity Fusion and Zero Matrix Injection}

The usage of GS$^2$-RS is shown in Figure \ref{SSRS-5}. With four well-trained generators $\mathcal{G}_\text{r}^{\text{in}}$, $\mathcal{G}_\text{f}^{\text{in}}$, $\mathcal{G}_\text{r}^{\text{sa}}$, $\mathcal{G}_\text{f}^{\text{sa}}$, we feed objective user $u$'s condition vectors $\textbf{c}_u^{\text{in}}$, $\textbf{c}_u^{\text{sa}}$, and noise vectors $\textbf{z}_u^{\text{in}}$, $\textbf{z}_u^{\text{sa}}$, we can achieve the synthetic vectors $\overline {\bf{r}} _u^{{\rm{in}}}$, $\overline {\bf{f}}_u^{{\rm{in}}}$, $\overline {\bf{r}} _u^{{\rm{sa}}}$, $\overline {\bf{f}}_u^{{\rm{sa}}}$, which are formulated as follows: 

\begin{equation}
	\begin{aligned}
	\mathcal{G}_{u/{\rm{f}}}^*({\bf{z}}_u^*,{\bf{c}}_u^*){\rm{ =  < }}\overline {\bf{r}} _u^*,\overline {\bf{f}} _u^* > ,
	\end{aligned}
	\label{eq4} 
\end{equation}
where $^*$ denotes either $^{\text{in}}$ or $^{\text{sa}}$. For each objective user, our propose model can generate several synthetic ${\overline {\rm{\textbf{r}}} _u^*}{\rm{,}}{\overline {\rm{\textbf{f}}} _u^*}$ pairs. Generally, we treat each vector pairs as a virtual user $u'$ for the objective user $u$, named \textit{self neighbors}. And each virtual user $u'$'s preferences can be formulated as follows:
\begin{equation}
{\bf{r}}_{u'}^* = {\bf{\bar r}}_u^* \odot {\bf{\bar f}}_u^{\rm{*}}.
	\label{eq5} 
\end{equation}

Note that the number of virtual users can be tuned for better performance. We will discuss it in our experiment section. For the sake of simplicity, we only generate $t$=2 self neighbors here for explanations, as shown in Figure \ref{SSRS-5}. Then we feed the virtual user $u'$'s preference vectors and original user $u$'s preference vector $\textbf{r}^*$ into existing CF models to achieve self preference ${\bf{\bar r}}^*_u$ (Formula \ref{eq7}), for interest and satisfaction, respectively. This operation can generate two enhanced preference matrices ${\bar R^{\text{in}}}$, ${\bar R^{\text{sa}}}$. Note that we can achieve these two matrices with different operations among original vectors and their virtual neighbors, such as average, threshold mechanism, collaborative filtering, etc.
\begin{equation}
{\bf{\bar r}}^*_u= \text{CFmodel}({\textbf{r}^*_u},\textbf{r}{^*_{u'1}},...\textbf{r}{^*_{u't}}).
\label{eq7} 
\end{equation}

Then serendipity fusion operation is employed for marking each potential candidate item for objective users. First, we give a reminder of \textit{serendipity items}: the items with high relevance but low shopping purpose \cite{A16DBLP:journals/tmm/YangXWHY18}. With this consideration, intuitively, we can deduce that the items with high satisfaction but low interest should be marked as serendipity items, which means that the objective user would achieve a much better experience after buying them. To achieve the serendipity item set, we set the thresholds $\theta ^*$ for interest and satisfaction, respectively. The item with $r^{\text{sa}} \geq \theta^{\text{in}}$ but $r^{\text{in}} < \theta^{\text{in}}$ is marked as $s_{ui}$=1. For each user, there is an indicator $\textbf{s}_u$ to mark his serendipity items.

One of our contributions is to solve the cold-start problem. The reason for the cold-start issue is the sparsity of the user-item matrix. Thus we employ a zero matrix injection to relieve the sparsity issue. Unlike existing matrix injection methods, we do not inject the potential candidate items into the user-item matrix. Instead, we pick impossible items to filter the candidate items. The reason is that 1) the rating deduction from users' preferences is usually tricky and inaccurate, with many uncontrollable factors. 2) In real-world scenarios, the potential items of an objective user take a relatively small partition from thousands or millions of unobserved items. So adding zeros for impossible items could relieve the sparsity issue greatly.

While we filter the items with two vectors ${{\bf{\bar r}}^{{\text{in }}}_u}$ and ${{\bf{\bar r}}^{{\text{sa}}}_u}$ for user $u$'s zero injections, there are several situations with different values (${{\bar r}^{{\text{in }}}_{ui}}$ and ${{\bar r}^{{\text{sa}}}_{ui}}$, $i$-th is the location index of both vectors), respectively. Note that the ${{\bf{\bar r}}^{{\text{in }}}_u}$ and ${{\bf{\bar r}}^{{\text{sa}}}_u}$ are computed by Formula \ref{eq5}, and the element value is [$\bar r^*                                                                                                                                                                                                      _{ui}$, 0, ?], $0 < \bar r_{ui} \leq 1$. Generally, we obey the following principles to inject $r^\text{h}_{ui}=0$ into enhanced matrix $\textbf{R}^\text{h}$:
\begin{itemize}
	\item If ${{\bar r}^{{\text{in}}}_{ui}}<\theta^{\text{in}}$ and ${{\bar r}^{{\text{sa}}}_{ui}}<\theta^{\text{sa}}$, inject $r^\text{h}_{ui}=0$;
	\item If ${{\bar r}^{{\text{in}}}_{ui}}<\theta^{\text{in}}$/${{\bar r}^{{\text{sa}}}_{ui}}<\theta^{\text{sa}}$, and ${{\bar r}^{{\text{sa}}}_{ui}}/{{\bar r}^{{\text{in}}}_{ui}}=?$, inject $r^\text{h}_{ui}=0$;
	\item Else, we set $r^\text{h}_{ui}=r_{ui}$.
\end{itemize}

Intuitively, the impossible item set for recommendations consists of low interest and low satisfaction (below the thresholds) items. Also, we inject 0s for the items with an unknown preference (?) and a low interest/satisfaction preference. As far as we know, these 0s can relieve the user-item matrix's sparsity. Moreover, the enhanced user-item matrix $\textbf{R}^\text{h}$ can give more meaningful feedback to indicate users' preferences on items and reduce the unknown feedbacks, which relieves the cold-start problem. While for the items/users with few or zero feedbacks (new items/users in recommender systems) in original user-item matrix $\textbf{R}$, our proposed model can inject some zeros as an initialization. It relieves the new user/item cold-start problem. Note that these zeros are temporary, not fixed. Once the original user-item matrix has been updated to a large extent (utilizes many new $r_{ui}$ to replace ?), We can employ GS$^2$-RS again to update $\textbf{R}^\text{h}$.

With $\textbf{R}^\text{h}$ as input (replacing original \textbf{R}), and interest matrix ${{\bar {\textbf{R}}}}^{\text{in}}$, satisfaction matrix ${\bar {\textbf{R}}}^{\text{sa}}$ and self-serendipity matrix \textbf{S} as side information inputs, we can achieve different types (Top-\textit{k}, CTR, or next purchase, etc) recommendation results \textbf{L} with different RS models:
\begin{equation}
\bf{L}{\rm{ = RS model ({{\textbf{R}}}^\text{h},{\bar {\textbf{R}}}^{\text{in}},{\bar {\textbf{R}}}^{\text{sa}}, \textbf{S})}}.
\label{eq6} 
\end{equation}
The complete GS$^2$-RS model is described in Algorithm 1:
 \begin{algorithm}[!h]
 	\caption{Generative Self-Serendipity RS model}
 	\begin{algorithmic}[1]
 		\REQUIRE Original user-item matrix $\textbf{R}$.
 		\ENSURE Recommendations results $\textbf{L}$.\\
 		\STATE Initializations: condition vectors {$\textbf{c}^*$}, parameters and thresholds.\\
 		\textbf{Step 1: Preference Modelling}
 		\STATE Calculate interest matrix $\textbf{R}^{\text{in}}$, satisfaction matrix $\textbf{R}^{\text{sa}}$ with threshold $\alpha_{ui}$ (Figure \ref{SSRS-3});
 		\STATE Train CGANs with Formula \ref{eq2} and Formula \ref{eq3};
 		\STATE Output generators $\mathcal{G}_\text{r}^{\text{in}}$, $\mathcal{G}_\text{f}^{\text{in}}$, $\mathcal{G}_\text{r}^{\text{sa}}$, $\mathcal{G}_\text{f}^{\text{sa}}$;\\
 		\textbf{Step 2: Self-serendipity Fusion\&Zero Injection}
 		\FOR{objective user $u$} 
 		\STATE Ensure self-neigbor number $t$;
 		\FOR{each self-neigbor $u'$} 
 		\STATE Generate $\overline {\bf{r}} _u^{{\rm{in}}}$, $\overline {\bf{f}}_u^{{\rm{in}}}$, $\overline {\bf{r}} _u^{{\rm{sa}}}$, $\overline {\bf{f}}_u^{{\rm{sa}}}$ with $\mathcal{G}_\text{r}^{\text{in}}$, $\mathcal{G}_\text{f}^{\text{in}}$, $\mathcal{G}_\text{r}^{\text{sa}}$, $\mathcal{G}_\text{f}^{\text{sa}}$;
 		\STATE Calculate $\textbf{r}^{\text{in}}_{u'}$, $\textbf{r}^{\text{sa}}_{u'}$ with Formula \ref{eq5};
 		\ENDFOR
 		\STATE Calculate ${\bf{\bar r}}^\text{in}_u$, ${\bf{\bar r}}^\text{sa}_u$ with Formula \ref{eq7}; 
 		\STATE Compare ${\bf{\bar r}}^\text{in}_u$ with ${\bf{\bar r}}^\text{sa}_u$ element-wised with thresholds $\theta^{\text{in}}, \theta^{\text{sa}}$;
 		\STATE Mark the self-serendipity indicator $s_{ui}$;
 		\STATE Inject 0s to form \textbf{r}$^\text{h}_{ui}$;
 		\ENDFOR
 		\STATE Output $\textbf{R}^\text{h}$, ${\bar {\textbf{R}}}^{\text{in}}$, ${\bar {\textbf{R}}}^{\text{sa}}$ and \textbf{S};\\
 		\textbf{Step 3: Recommendations}
 		\STATE Input $\textbf{R}^\text{h}$ instead of \textbf{R} into SOTA RS models;
 		\STATE Input ${\bar {\textbf{R}}}^{\text{in}}$, ${\bar {\textbf{R}}}^{\text{sa}}$ and \textbf{S} as side information into SOTA RS models;
 		\STATE Calculate \textbf{L} with Formula \ref{eq6};
 		\RETURN Recommendations results $\textbf{L}$.
 	\end{algorithmic}
 \end{algorithm}
\subsection{Recommendation Analysis}
This section introduces how our proposed model enhances the SOTA recommender systems and solves the filter-bubble problem.
\subsubsection{Enhancing CF/MF/NN based Recommenders}
Existing recommender systems usually consist of three specific categories: collaborative filtering-based RS (CF models), matrix factorization-based RS (MF models), and neural network-based RS (NN models). GS$^2$-RS can enhance both recommender systems for the following reasons: For enhancing CF models, GS$^2$-RS could give more valuable feedbacks in $\textbf{R}^\text{h}$, (0s that injected in Zero Matrix Injection), which can help the model to compute the distance between different users/items, and select more accurate users' to filter the items. 

For MF models, we employ WRMF or other matrix factorization to learn users'/items' latent vectors from user-item matrix \textbf{R}, and then make recommendations. Note that all the existing matrix factorization algorithms' performance is affected dramatically by the matrix's sparsity, while our proposed model GS$^2$-RS can relieve the sparsity problem by replacing \textbf{R} with $\textbf{R}^\text{h}$. Moreover, the 0s in $\textbf{R}^\text{h}$ also can be treated as ratings, which can restrict the learned latent users'/items' vectors for accurate performance. Note that MF models are usually employed as the preprocessing for NN models to get the input of the neural network framework. Moreover, the input is vital for NN models. Indeed, our proposed model can enhance NN models by offering a superior input. We will give a discussion in the experimental section.                          
\subsubsection{Enhancing Personalized Recommendations for the Filter-Bubble problem}
Personalized Recommendation is an important factor in recommender systems because a boring, homogenous recommendation is not expected for each individual. For enhancing personalized recommendations, GS$^2$-RS employs users’ preferences (interest, satisfactions, and self-serendipity) to decide the recommendation orders in the recommendation results \textbf{L}. With interest matrix ${\bar {\textbf{R}}}^{\text{in}}$, satisfaction matrix ${\bar {\textbf{R}}}^{\text{sa}}$ and self-serendipity matrix \textbf{S}, we can make a fine-grained user profiling for personalized recommendations. For example, we could check the percentage of interest and satisfaction items in historical records for an objective user to distinguish the user's distribution on preferences. Then we rerank the recommendation list in accord with this preference distribution. Moreover, the self-serendipity item should also be considered for achieving personalized, accurate recommendations. Details are introduced in the experimental section. With these side information, the filter-bubble problem can be adequately solved.  
\section{Experiments}
This section validates our proposed framework with three aspects: 1) How GS$^2$-RS enhances the overall recommendation performance. 2) How GS$^2$-RS solves the cold-start and filter-bubble problems, and 3) How the threshold affects GS$^2$-RS's performance.
\subsubsection{Datasets}
We utilize two publicly accessible datasets: Movielens\footnote{http://grouplens.org/datasets/movielens/} and Amazon\footnote{http://www/kaggle.com/snap/amazon-fine-food-reviews/}. Detailds are indicated in Table \ref{AT1}. Grid search and 5-
fold cross validation are used to find the best parameters. In our proposed GS$^2$-RS, the thresholds are set: ${\alpha _u} = \sum\limits_i {{r_{ui}}} /\# num {(r_{ui})}$, $\theta^{\text{in}}=\theta^{\text{sa}}=0.5$. The learning rate is 0.01.
\begin{table}[htbp]
	\centering
	\caption{Datasets Statistics}
	\begin{tabular}{|c|c|c|c|c|}
		\hline
		Datasets & \#\textit{Users} & \#\textit{Items} & \#\textit{Feedbacks} & Sparsity\\
		\hline
		Movielens & 6,040 & 3,952 & 1,000,209 & 95.81$\%$\\
		\hline
		Amazon  & 16,619 & 37,762 &  256,287 & 99.95$\%$\\
		\hline
	\end{tabular}%
	\label{AT1}%
\end{table}
\subsubsection{Baselines}
To validate GS$^2$-RS,  we select several classic RS models and SOTA RS models as the validations: \textit{1) Collaborative Filtering} (CF) \cite{A21DBLP:reference/sp/KorenB15} and
\textit{2) Weighted Matrix Factorization} (WMF) \cite{A22DBLP:conf/cikm/KoenigsteinRS12,A24DBLP:conf/aaai/Chen0ZSCF020} are widely applied RS models, and \textit{3) Neural Collaborative Filtering} (NCF) \cite{A12DBLP:conf/www/HeLZNHC17}: a general neural network based recommendation framework, which emppoys GMF as its preprocessing.  \textit{4) Joint Variational Autoencoder} (JoVA) \cite{A23DBLP:conf/sigir/AskariSS21}: an ensemble of two VAEs to capture user-user and item-item correlations simultaneously for recommendation. \textit{5) Augmented Reality CF} (AR-CF) \cite{A3DBLP:conf/sigir/ChaeKCK20}: a GAN based CF model, which is applied directly on ratings.
\subsubsection{Metrics}
We employ standard metrics to validate overall recommendation performance, including Precision, Recall, and Normalized Discounted Cumulative Gain (NDCG), and Mean Reciprocal Rank (MRR).

For cold-start problem effectiveness, we utilize Exposure Ratio (ER) as the metric. Formally, the exposure ratio is computed
by B/A where B is the number of cold-start items which are exposed
to at least one user, and A is the number of the entire cold-start
items. For filter-bubble problem effectiveness, we utilize diversity (DI) and serendipity (SE) \cite{A16DBLP:journals/tmm/YangXWHY18} as metrics. Note that both metrics should be applied on a recommendation item set, as follows ($\textbf{I}^{\text{real}}$ is the ground truth):

${\rm{DI}} = \# categorynum({{\bf{I}}^{{\rm{rec}}}})/\# num({{\bf{I}}^{{\rm{rec}}}})$; 

${\rm{SE}} = \# num({{\bf{I}}^{{\rm{rec}}}} \cap {{\bf{I}}^{{\rm{real}}}} \cap ({{\bf{I}}^{{\rm{sa}}}} - {{\bf{I}}^{{\rm{in}}}}))/\# num({{\bf{I}}^{{\rm{real}}}})$.
\subsection{Overall Performance for Enhancing Recommendation}
The overall performance across two datasets is shown in Table \ref{AT2}. Generally speaking, GS$^2$-RS outperforms all compared models on both datasets for all metrics. The improvement can be attributed to two aspects: 1) The improvement from two GAN-based models (AR-CF and GS$^2$-RS) over other models indicates that the usage of GAN on sparse matrices is obvious for recommendations. 2) GS$^2$-RS can enhance performance by generating their virtual preferences, which is effective than directly generating their virtual ratings on items, which can be indicated from GS$^2$-RS's improvement on AR-CF. 
\begin{figure}[t] \centering  
	\subfigure[\textit{Pre.} on Movielens] { \label{fig:subfig:f1}     
		\includegraphics[width=0.48\columnwidth]{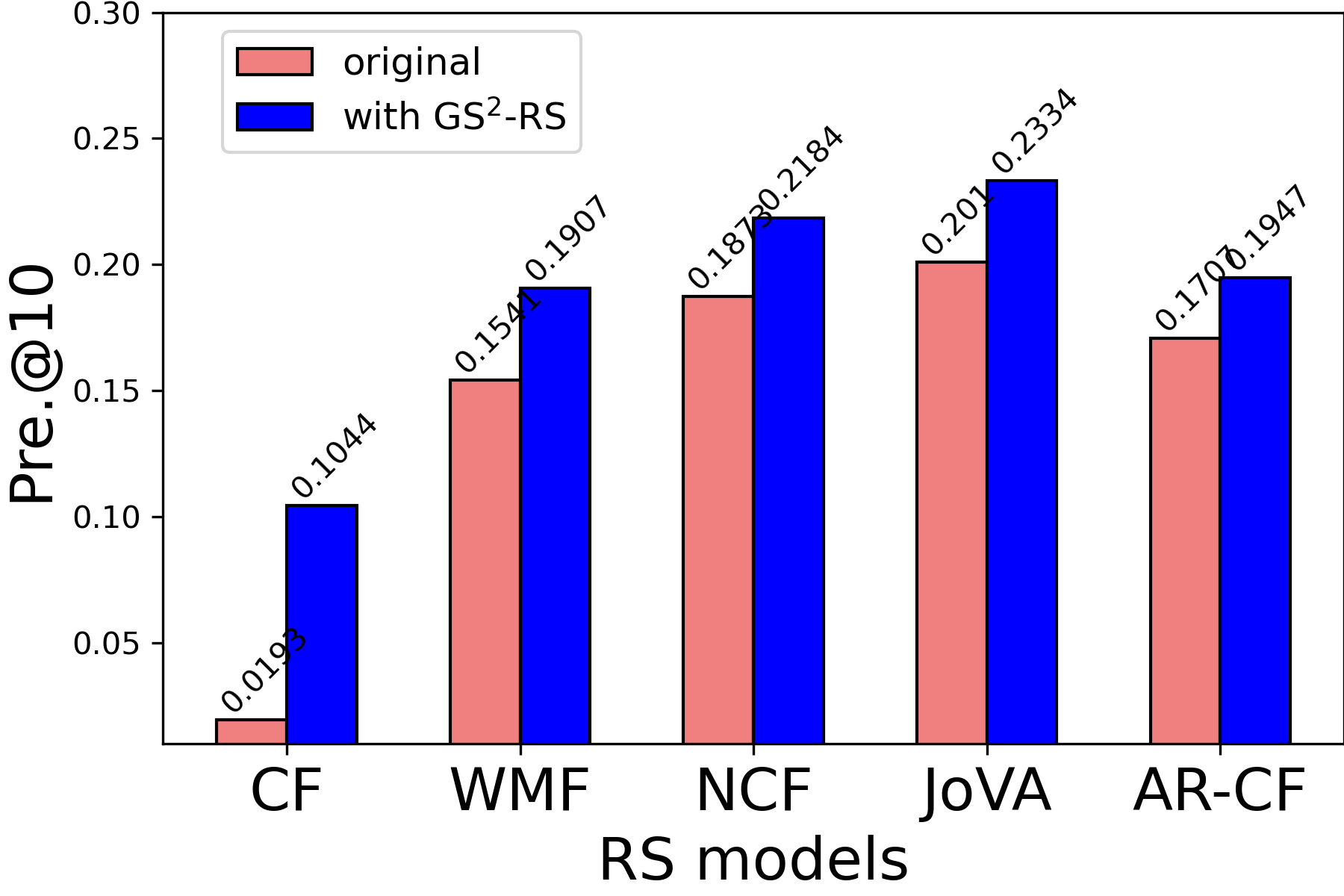}  
	}     
	\hspace{-9pt}
	\subfigure[\textit{NDCG} on Movielens] { \label{fig:subfig:f2}     
		\includegraphics[width=0.48\columnwidth]{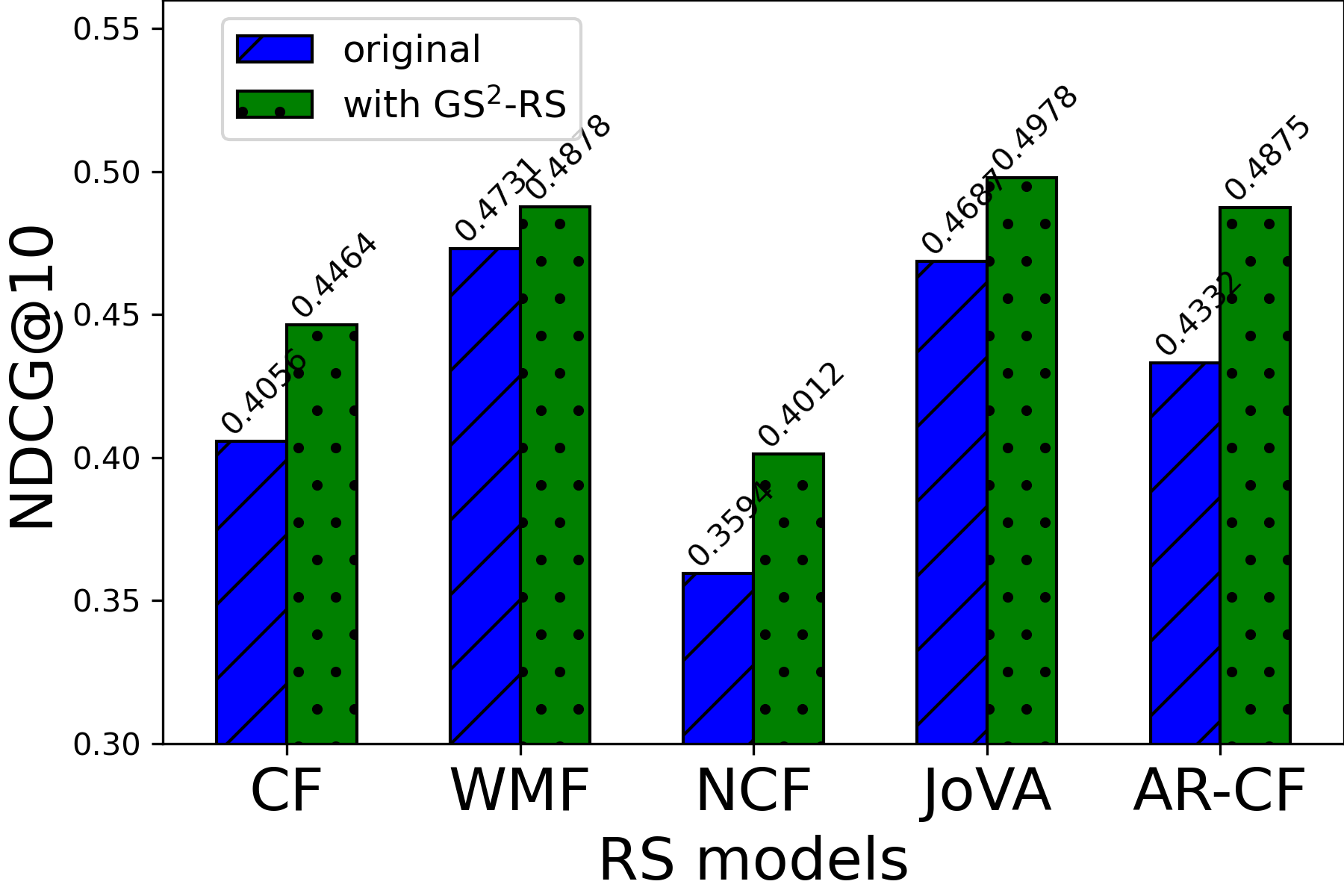}    
	}      
	\caption{Enhancing RS models as Preprocessing.}     
	\label{SSRS-6}     
\end{figure}
\begin{figure}[t] \centering  
	\subfigure[Movielens] { \label{fig:subfig:f3}     
		\includegraphics[width=0.48\columnwidth]{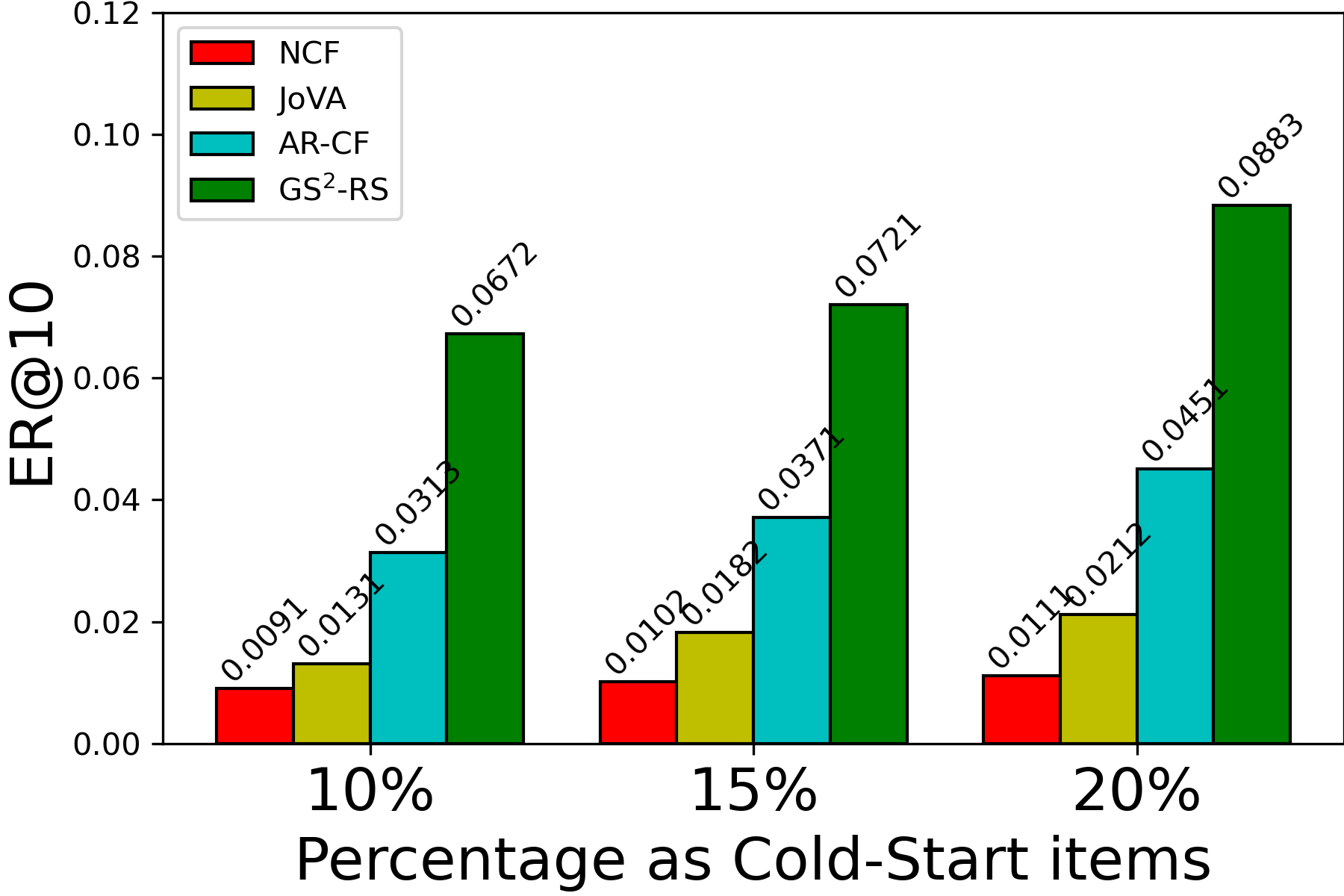}  
	}     
	\hspace{-9pt}
	\subfigure[Amazon] { \label{fig:subfig:f4}     
		\includegraphics[width=0.48\columnwidth]{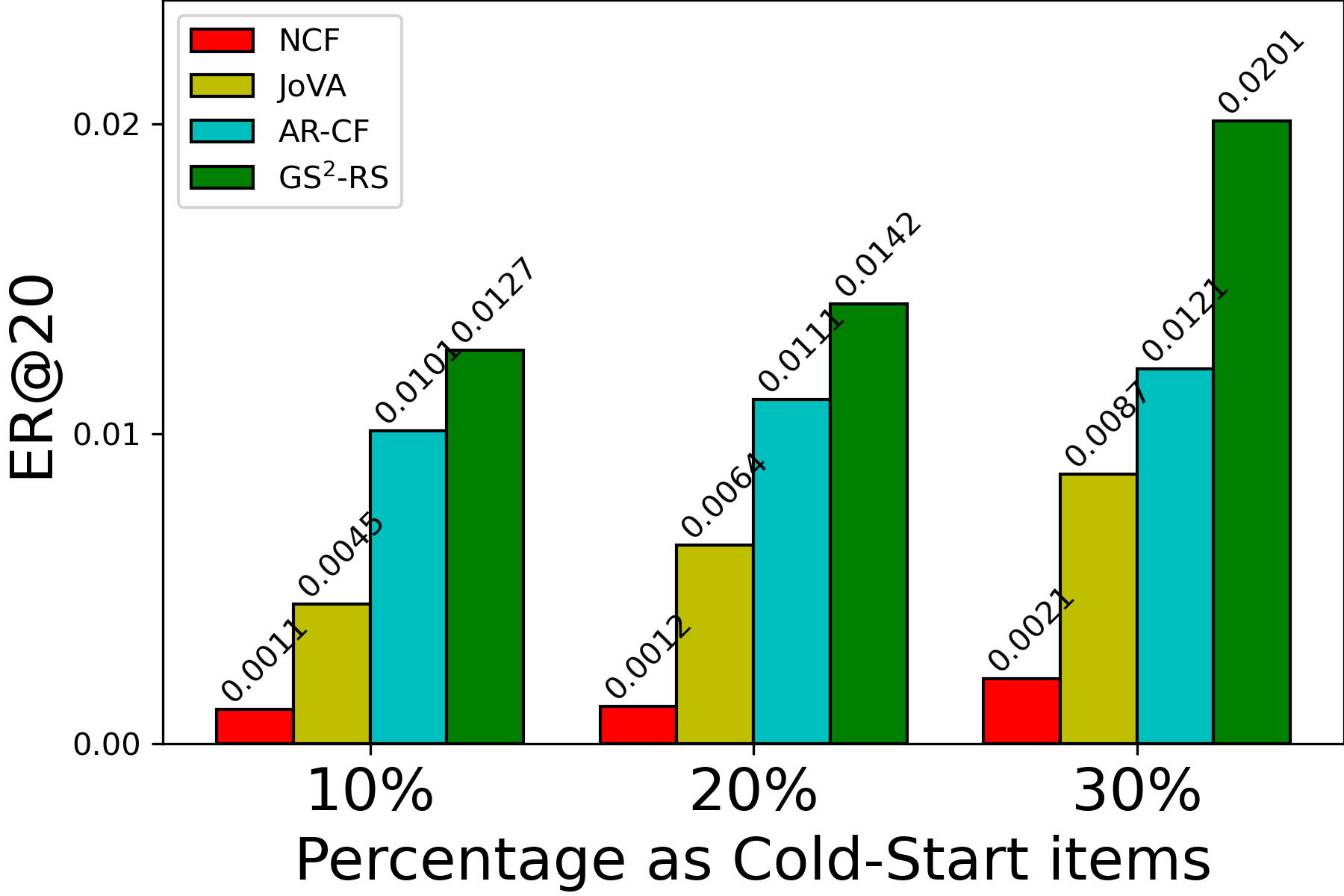}    
	}      
	\caption{Exposure ratio for cold-start items.}     
	\label{SSRS-7}     
\end{figure}

Moreover, as we claimed, GS$^2$-RS can be applied as the preprocessing for SOTA RS models, which are shown in Figure \ref{SSRS-6}. We observe that our proposed model can enhance the SOTA models' Precision and NDCG performance on Movielens. GS$^2$-RS universally and consistently provide the best accuracy, and we believe that the benefits are credited to GS$^2$-RS’s
characteristics that it can take advantage of the performance
gains coming from the generated virtual (but plausible) users' preferences as qualified training data. 
\begin{table*}[htbp]
	\centering
	\caption{The overall performance metrics of the compared methods for recommendations. The boldface font denotes the winner and $^*$ denotes the second winner in that columen. \textit{Perform.}+ denotes the preformance gain percentage on second best model. And the improvement is significant with t-test \textit{p}$<$0.05.}
	\begin{tabular}{|c|c|c|c|c||c|c|c|c|}
		\hline
		Datasets & \multicolumn{3}{c}{Movielens}&& \multicolumn{3}{c}{Amazon}&\\
		\hline
		Metrics & \textit{Pre.}@10 & \textit{Rec.}@10 & \textit{NDCG}@10 & \textit{MRR}& \textit{Pre.}@10 & \textit{Rec.}@10 & \textit{NDCG}@10 & \textit{MRR}\\
		\hline
		CF & 0.0193 & 0.0893 & 0.2210 & 0.4056 & 0.0088 & 0.0049 & 0.1129 & 0.3656\\
		\hline
		WMF  & 0.1541 & 0.1644 &  0.3059 & 0.4731$^*$ & 0.1032 & 0.1321 & 0.2048 & 0.4111\\
		\hline
		NCF  & 0.1873 & 0.1831 &  0.2710 & 0.3594 & 0.1321 & 0.1224 & 0.2321 & 0.3321\\
		\hline
		JoVA  & 0.2010$^*$ & 0.2001$^*$ &  0.3010 & 0.4687 & 0.1177 & 0.1331 & 0.3015 & 0.4014\\
		\hline
		AR-CF  & 0.1707 & 0.1127 &  0.3971$^*$ & 0.4332& 0.1421$^*$ & 0.1644$^*$ & 0.4015$^*$ & 0.4233$^*$\\
		\hline
		GS$^2$-RS  & \textbf{0.2230} & \textbf{0.2006} &  \textbf{0.4449} & \textbf{0.4837}& \textbf{0.1534} & \textbf{0.1756} & \textbf{0.4333} & \textbf{0.5210}\\
		\hline	
		\textit{Perform.}+  & \textit{10.94\% }& \textit{0.24\% } &  \textit{12.03\%}  & \textit{2.24\%} & \textit{7.95\%}  & \textit{6.81\%}& \textit{7.92\%} & \textit{23.08\%}\\
		\hline
	\end{tabular}%
	\label{AT2}%
\end{table*}
\subsection{Performance for Solving Cold-Start Problem}
The problem with the cold-start items is that they are challenging to be recommended. So we employ Exposure Ratio (ER) to evaluate the performance for solving the cold-start problem. We employ NCF, JoVA, AR-CF as baselines compared with GS$^2$-RS, as shown in Figure \ref{SSRS-7}. \textit{T}$\%$denotes the bottom percentage of items that have interactions with users, and we treat these items as cold-start items. We observe that NCF and JoVA have difficulty solving the cold-start problem, and AR-CF and our model outperform them considerably. Meanwhile, because our proposed model generates users' preferences, not ratings, GS$^2$-RS performs better than AR-CF. Jointly considered with accuracy results in Figure \ref{SSRS-6}, the results demonstrate the effectiveness of our model in solving cold-start issues with a stable recommendation performance.  
\subsection{Performance for Solving Filter-Bubble Problem}
\begin{figure}[t] \centering  
	\subfigure[Diversity (DI) @10] { \label{fig:subfig:f5}     
		\includegraphics[width=0.48\columnwidth]{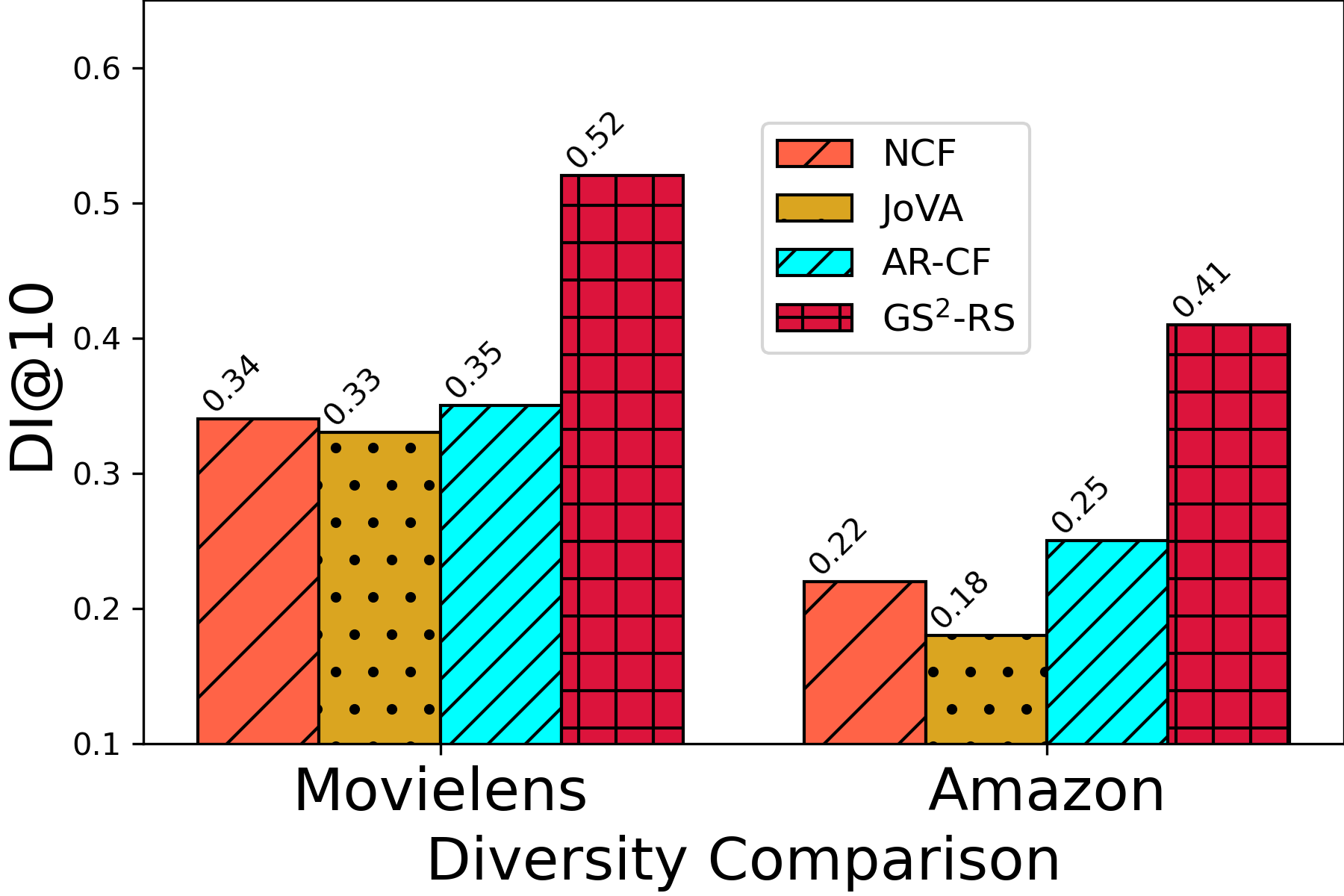}  
	}     
	\hspace{-9pt}
	\subfigure[Serendipity (SE) @10] { \label{fig:subfig:f6}     
		\includegraphics[width=0.48\columnwidth]{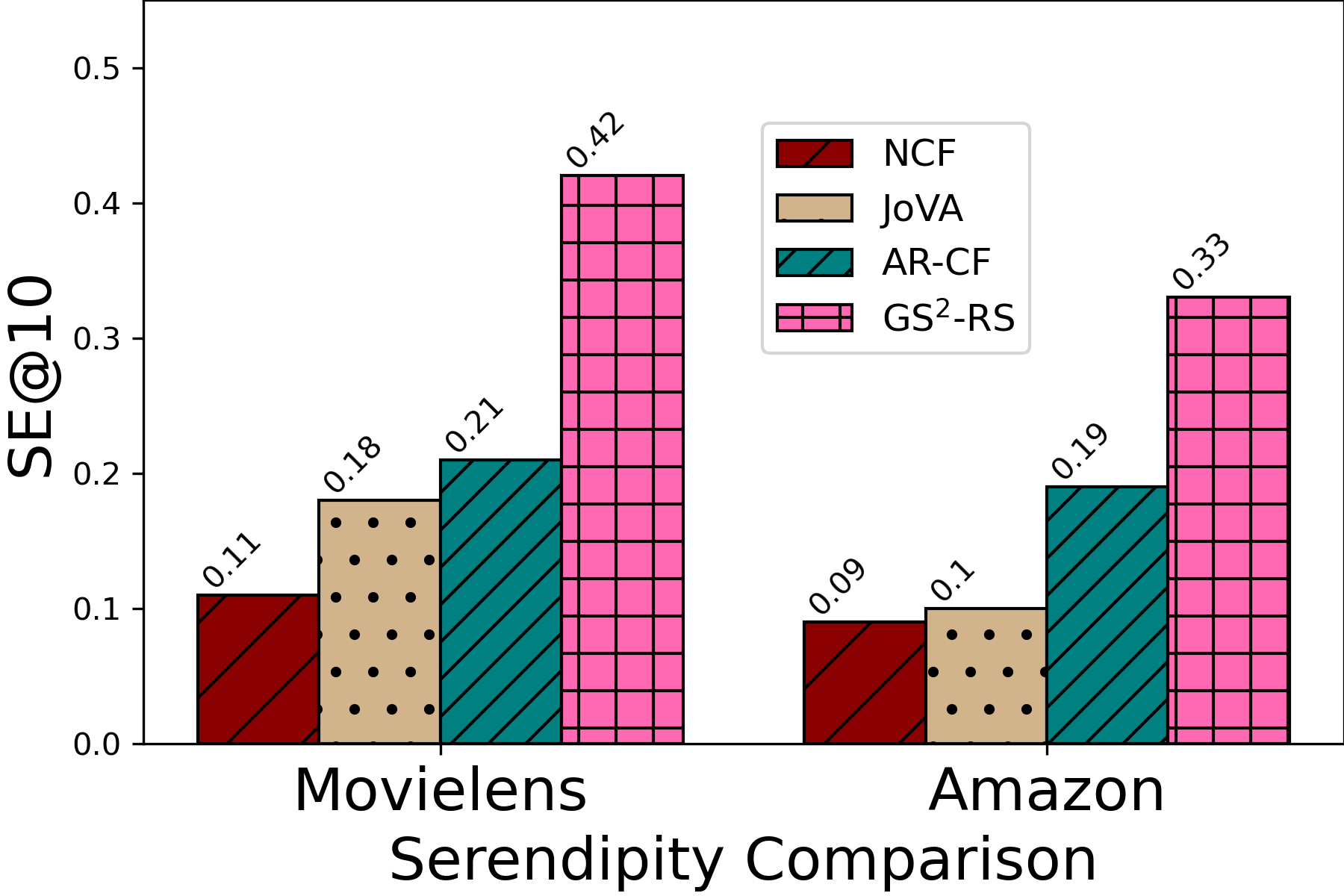}    
	}      
	\caption{Diversity/Serendipity for the FB problem.}     
	\label{SSRS-8}     
\end{figure}

We utilize Diversity and Serendipity to evaluate the performance for solving the filter-bubble problem, as shown in Figure \ref{SSRS-8}. From the results, we observe that 1) GS$^2$-RS can improve the diversity and serendipity significantly, especially on sparse dataset Amazon. 2) JoVA achieves a second-best accuracy performance but a relatively low diversity and serendipity compared with other baselines. With a high diversity and serendipity, we can offer users an attractive and exciting recommendation instead of a boring, repeated one, greatly relieving the filter-bubble problem.
 \begin{figure}[tbp]
	\centering
	\includegraphics[width=1\columnwidth]{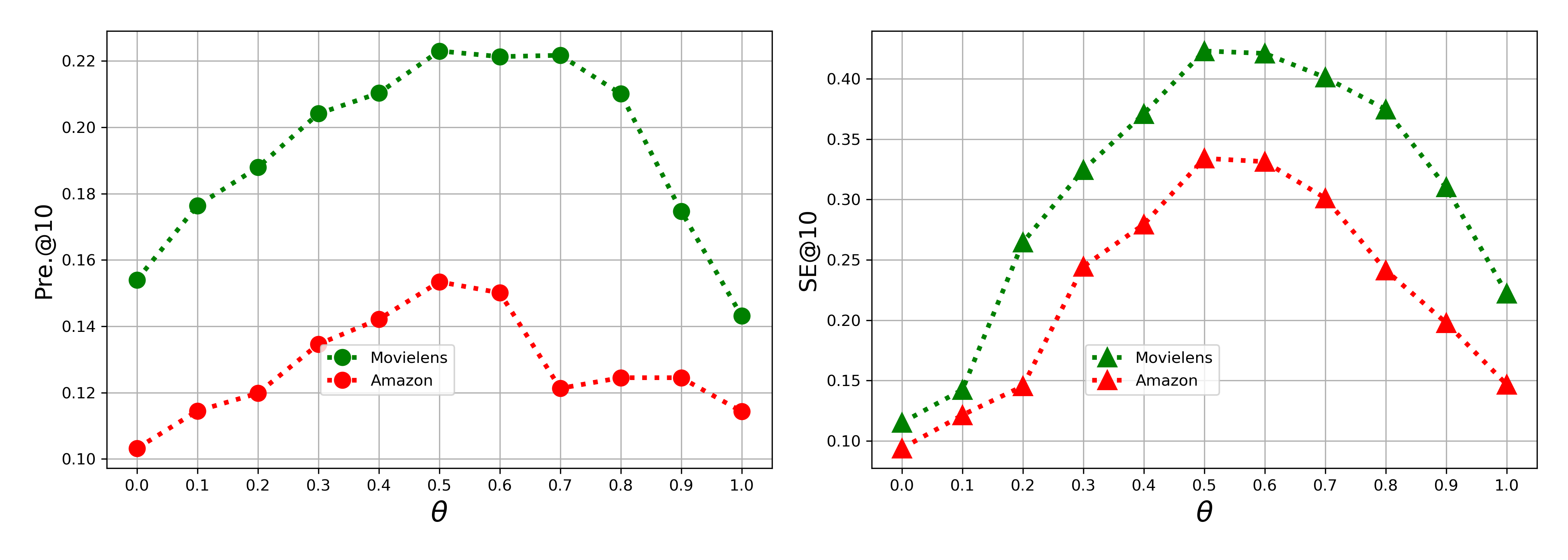}
	\caption{Effect of threshold $\theta$ on accuracy and serendipity.}
	\vspace{-10pt}
	\label{SSRS-9}
\end{figure}
\subsection{Threshold Effect Analysis}
We validate the effect of $\theta$, the most vital threshold of GS$^2$-RS, ranging (0.0, 1,0) with step 0.1. Note that we set $\theta^{\text{in}}$=$\theta^{\text{sa}}$=$\theta$ for validation, as shown in Figure \ref{SSRS-9}. From the results, we observe that GS$^2$-RS achieves the best performance at $\theta$=0.5. When $\theta$ is descending to 0, GS$^2$-RS can not filter any items, which fades to a basic GAN-based model. When $\theta$ is ascending to 1, GS$^2$-RS drops every item and damages its performance.  
\section{Related Work}
As well-known basic problems in the recommendation, cold-start and filter-bubble problems are explored by many researchers recently: For the cold-start problem, \cite{A10DBLP:conf/sigir/ChenDHW20} proposed a tagging algorithm to tag unobserved items for relieving cold-start issue. \cite{A9DBLP:conf/sigir/BiSYWWX20a} utilized cross-domain information to reduce data sparsity for cold-start problem and achieved the SOTA recommendation performance. Meanwhile, some researchers explore the usage of GAN \cite{A19DBLP:conf/nips/GoodfellowPMXWOCB14} for the cold-start problem: \cite{A3DBLP:conf/sigir/ChaeKCK20} generated virtual neighbor for objective users and made accurate recommendations by reducing cold-start items. \cite{A18DBLP:conf/aaai/Wang21a} generated users' embedding for recommendation and improved \textit{NDCG}@100 significantly. However, these frameworks do not consider generating users' preferences, especially fine-grained preferences (as our framework GS$^2$-RS).

For the filter-bubble problems, some researchers \cite{A11DBLP:conf/recsys/BurbachNPZV18,A21DBLP:reference/sp/KorenB15} tried to add the diversity of recommendation list to tackle it. \cite{A6DBLP:conf/recsys/KapoorKTKS15} adapted users' novelty preferences into recommendations, which added the diversity for relieving the filter-bubble problem. Recently, the idea of serendipity has been proposed to solve the filter-bubble problem by offering novel, diverse and high-satisfaction recommendations. \cite{A17DBLP:journals/jcst/ZiaraniR21} gave general explanations about why serendipity items could work for tackling the filter-bubble situations. \cite{A16DBLP:journals/tmm/YangXWHY18} proposed a matrix factorization-based model for enhancing serendipity for superior recommendations. However, the challenge is utilizing serendipity into the recommender system framework to solve cold-start and filter-bubble problems simultaneously.
\section{Concluding Remarks}
We have introduced GS$^2$-RS, a novel framework for addressing cold-start and filter-bubble problems with the CGAN framework and matrix injection method. From empirical experiments on public datasets, we demonstrated that GS$^2$-RS is effective in dealing with both problems: GS$^2$-RS significantly advances accuracy, diversity, and serendipity compared to SOTA RS models. In the future, we plan to explore extending GS$^2$-RS to incorporate images (with GCN), social networks (with KG), or context (with NLP).
\bibliography{self-cf}
\end{document}